\documentclass[journal=aaembp, manuscript=article]{achemso}

\usepackage{graphicx}
\usepackage{float}
\usepackage[T1]{fontenc}
\graphicspath{./}

\usepackage{hyperref}

\author{Mohammed Ghadiyali}
\email{ghadiyali.mohd@physics.mu.ac.in}
\affiliation{Department of Physics, University of Mumbai, Kalina Campus,
	Santacruz (E), Mumbai - 400 098, India.}
\author{Sajeev Chacko*}
\email{sajeev.chacko@physics.mu.ac.in; sajeev.chacko@gmail.com}
\affiliation{Department of Physics, University of Mumbai,
			Kalina Campus, Santacruz (E), Mumbai - 400 098, India.}

\def  \degree  {{$^\circ$}}

\title{Hydrogenated-Graphene encapsulated Graphene: A versatile material for device applications}

\abbreviations{GR, SOC, QSHI, QSHE, 2D-BN, HG, GR-BN, DFT, QE, ATK, PDOS, SI}

\begin{document}
\begin{abstract}
Graphene and its heterostructures exhibit interesting electronic properties and are explored 
for quantum spin Hall effect(QSHE) and magnetism based device-applications.
In present work, we propose a heterostructure of graphene encapsulated by hydrogenated-graphene 
which could be a promising candidate for a variety of device applications.
We have carried out DFT calculations on this system to check its feasibility to be a versatile 
material.
We found that electronic states of multilayer pristine graphene, especially Dirac cone, an 
important feature to host QSHE, can be preserved by sandwiching it by fully 
hydrogenated-graphene.
Interference of electronic states of hydrogenated-graphene was insignificant with those of 
graphene.
States of graphene were also found to be stable upon application of electric field up-to 
2.5V/nm.
For device applications, multilayer-graphene or its heterostructures are required to be 
deposited on a substrate, which interacts with system opening up a gap at Dirac cone making it 
less suitable for QSHE applications and hydrogenated-graphene can prevent it.
Magnetization in these hydrogenated-graphene sandwiched graphene may be induced by creating 
vacancies or distortions in hydrogenated-graphene, which was found to have minimal effect on 
graphene’s electronic states, thus providing an additional degree of manipulation.
We also performed a set of calculations to explore its applicability for detecting some 
molecules.
Our results on trilayer-graphene encapsulated by hydrogenated-graphene indicate that all these 
observations can be generalized to systems with a larger number of graphene layers, indicating 
that multilayer-graphene sandwiched between two hydrogenated-graphene is a versatile material 
that can be used in QSHE, sensor devices, etc.
\end{abstract}

\maketitle


\section{Introduction}
\label{intoduction}

The honeycomb lattice structure of graphene (GR) is the nearest analogy to Haldane’s 
model~\cite{Haldane_Model} in condensed matter physics. 
However, due to the weak spin-orbital coupling (SOC) of carbon atoms, it is not an ideal 
quantum spin Hall insulator (QSHI) or 2D topological insulator~\cite{Graphene_Spin_Hall}.
Quantum spin Hall effect (QSHE) is realized in GR, but the band gap of GR is zero, limiting its 
application.
For this reason, a lot of work has been devoted to finding an alternative QSHI with high SOC 
and high bulk band gap.
A non-Haldane’smodel system HgTe quantum well has been found to be a good alternative.
However, it requires low operating temperature which hinders its 
applications~\cite{HgTe_Quantum_Well}.
Further, 2D allotropes of group VI(A) and V(A) have also been predicted to be 
QSHI~\cite{list_2d_material_review}.
Out of these, stanene (a 2D allotrope of tin) has been reported to meet most of the 
requirements and has been studied both theoretically and 
experimentally~\cite{Stanene_Synthesis_China, Stanene_Synthesis_China_2, 
	Stanene_Synthesis_India,Stanene_MG}.

Even after all these developments, GR, both in pristine as well as heterostructures forms are 
still being investigated for QSHE~\cite{Graphene-WS2_induce_SOC,  
	Bi-layer_Graphene_Topological_Transport}.
To preserve the QSHI nature, GR is generally coupled with 2D boron-nitrate (2D-BN) as a 
heterostructure, fabricated by alternating layers of GR and 2D BN~\cite{Grap-BN}.
Apart from 2D BN, other 2D materials like WS$_2$~\cite{Grap-WS2} and MoS${_2}$~\cite{Grap-DMC} 
- inducing spin-orbital coupling, yttrium iron garnet~\cite{Grap-YIG-1, Grap-YIG-2}, and 
EuS~\cite{Grap-EuS} - inducing ferromagnetism, MoSe${_2}$~\cite{Grap-MoSe2} - fabrication of 
Schottky devices, etc, have been studied for various applications.
One of the biggest drawbacks of these heterostructures is they strain GR, opening up a band gap 
in the place of Dirac cone and fabrication of these above-mentioned structures is not a trivial 
process, as explained further in the text.
In the present work, we propose an alternative to this system by replacing the 2D BN with 
hydrogenated-GR (HG).
This substitution preserves the essential Dirac cone of the GR layer enabling the composite 
multilayer system to host QSHE.
However, there are certain benefits to this approach over the GR-BN heterostructures as 
discussed below.

Mismatch in the lattices between the graphene and other layers may have profound effects on its 
structural and electronic properties.
For instance, a moiré pattern is seen in the crystal structure of the GR-BN heterostructures as 
a consequence of slight mismatch of their lattice vectors~\cite{moire-pattern-graphene}.
However, these are reported to form inside the GR and their effects on the QSHI states is 
required to be studied in detail.
Such an effect breaks the inversion symmetry of the GR opening up a band gap.
Further, the QSHI states in GR are reported to be quite fragile due to which high quality and a 
defect-free sample is required.
The preferred method has been to mechanically exfoliate the sheets of GR via scotch tape 
process~\cite{Grap-BN-Exp-1, Grap-BN-Exp-2}.
This process is also repeated for the fabrication of 2D BN sheets and then these sheets are 
stacked on each other in the required orientation.
Though this yields a high purity and defect-free sample, it requires a high level of skill and 
is time-consuming.
While a method like chemical vapor deposition is used, it creates nanoflakes of GR rather than 
a continuous film~\cite{Grap-BN-CVD}.
Other than that strain matching GR and 2D BN is again a complex process~\cite{Grap-BN-strain-1,
	Grap-BN-strain-2}. 
Hence, the large-scale extension of this process is challenging.

On the other hand, to fabricate the proposed systems one needs to exfoliate multilayer graphene 
and only hydrogenate the outermost layer.
A similar technique has been implemented for the fabrication of heterostructure of GR and 
monolayer transition-metal dichalcogenides~\cite{Hetro_fabrication}.
Or the recent developments in the fabrication of high-quality graphene can be used for the 
same~\cite{Graphene_synthesis}.
In addition, the magnetization in HG can be modulated via photolithographic 
processes~\cite{Mag-HG}.
Such modulations in the proposed GR-HG heterostructures can be used to tune its properties for 
desired magnetic applications.
We demonstrate this tunability for GR and HG trilayer nanoribbons.
Hence, the present work is divided into two parts: In the first part, we discuss the stability 
of the proposed systems and the role of HG in preserving the essential electronic states of the 
GR layer.
As the dimensions of the gate in nanodevices is being reduced, it can be described as 1D 
structures.
Hence, in the second part, we construct nanoribbons of up to 20 unit cells~\footnote{With the
	available computational facility we could do the calculations only for a maximum of 20 unit 
	cells.
	However, in order to observe QSHE a much wider nanoribbon of up to about 50 unit cells may 
	be required~\cite{Graphene_nanoribbon}} to study magnetization in it.
We also explore the ability of the HG-GR trilayer to adsorb simple molecules such as H$_2$, 
O$_2$, CO$_2$, and ethanol.

\section{Computational Details}
\label{CM}

We have performed the first principle density functional theory (DFT) calculations as 
implemented in the Quantum ESPRESSO (QE)~\cite{QE} package.
We have used Perdew-Burke-Ernzerhof (PBE)~\cite{PBE} exchange-correlation function and included 
the DFT-D2 van der Waals corrections~\cite{DFT-D_1, DFT-D_2} for representing the electron 
dispersion effect induced due to the multi-layer nature of the systems.
A $\Gamma$-centered Monkhorst-pack \textit{k}-point grid of $15\times15\times1$ gave was 
sufficient to sample the irreducible Brillouin zone.
A kinetic energy cutoff of 40 Ry for the wavefunctions gave the required convergence in the 
total energy.
This energy cutoff has been used in all the calculations presented in this work.
The pseudopotentials used in this work were obtained from the PSlibrary version 
1.0~\cite{PS_libary, PS_libary_2}.
Further, to limit the interaction between the periodic images of the unit cell, a vacuum 
of~$\approx10$\AA is added to the unit cell above and below the trilayer system.
The geometry optimization was performed until the forces per atom were reduced to less 
than~$10^{-6}$~Ha/au.
The phonon spectra calculations of the systems were performed by the density-functional 
perturbation method, as implemented in QE.

To compute the edge states, nanoribbons were constructed having a width of up to 20 unit cells 
along the y-direction and a single unit cell in the $X$-direction.
To minimize the interactions from neighboring images, a vacuum of $\approx10$\AA~was added on 
both sides along $Y$ and $Z$-directions.
The \textit{k}-point mesh of $6\times1\times1$ was used.
A small and arbitrary initial value of the magnetic moment was set for each atom for the 
spin-polarized calculations and a full geometry optimization was carry out to obtain the final 
magnetic state.
For computing the band structures, high symmetry \textit{k}-points of the Brillouin zone were 
selected along the path $M\rightarrow \Gamma\rightarrow K\rightarrow M$.
The \textit{k}-points were generated by the XCrysDens package~\cite{Xcrystal}.
It may be noted that we have not considered spin-orbit coupling since as it is quite weak in 
carbon and hydrogen atoms.

\begin{figure}[H]
	\centering
	\includegraphics[scale=0.75]{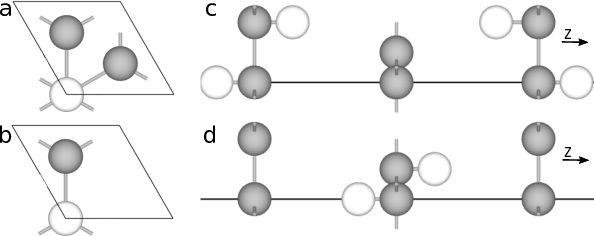}
	\caption{
		Configuration of trilayer HG-GR systems. (a) and (b) Top view the two different 
		stacking angles (0\degree and 60\degree, respectively), The axis of rotation is the 
		z-axis that is perpendicular to the plane of the paper. (c) and (d) Different stacking 
		orders (HG-GR-HG and GR-HG-GR, respectively). These structures are before geometry 
		optimization.
	}
	\label{complete_system}
\end{figure}

We studied two types of trilayer systems: (1) single GR layer sandwiched between two HG layers 
denoted as system A and (2) single HG layer sandwiched between two GR layers denoted as system 
B as illustrated in figures~\ref{complete_system}(c)and~\ref{complete_system}(d), respectively.
As these are multilayer systems, the stacking angle is also an important feature.
The two selected different stacking angles are 0\degree generally noted as AB configuration and 
60\degree generally denoted as AA configuration, as shown in figures~\ref{complete_system}(a) 
and~\ref{complete_system}(b), respectively.
The selection of the stacking angle was done in accordance with the Bernal stacking method.
For convenience, the systems have been named using the following nomenclature: a prefix is used 
to denote trilayer configuration and postfix to indicate the stacking angle.

As an example, the system of GR encapsulated by HG with stacking angle of 0\degree is named as 
``A0'', while the system where HG is encapsulated by GR with stacking angle of 60\degree is 
named as ``B60''.
Hence the structures have been named as A0, A60, B0, and B60.
Similar configurations for the bilayer systems were also studied.
However, from the phonon dispersion calculations, they were found to be very unstable and hence 
not considered for the presented work.
The crystal and band structure details regarding these additional systems can be found in the SI 
section VI.
The systems described in the text had been generated using ATK Virtual Nano Lab~\cite{VNL-2017}.

\section{Results and Discussion}

\subsection{QSHE in Graphene - HG heterostructures}

We first discuss the structures of the the GR-HG heterosystems and their stability.
We then describe by their band structures followed by the effect of electric field on the
electronic structure properties of these heterostructures.

\subsubsection{Structure and Stability of Proposed Systems}

\begin{figure}[H]
	\centering
	\includegraphics[scale=0.75]{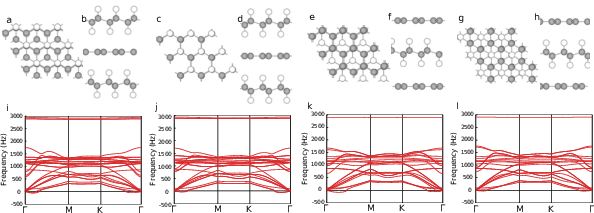}
	\caption{(color online)
		(a) Top and (b) side views of A0 system, (c) top and (d) side view of A60 system,
		(i), (j) phonon dispersion of the systems A0 and A60, respectively, plotted along the 
		high-symmetry \textit{k}-path,
		(e) Top and (f) side views of B0 system, (g) top and (h) side view of B60 system
		(i), (j) phonon dispersion of the systems B0 and B60, respectively, plotted along the 
		high-symmetry \textit{k}-path.
	}
	\label{fig:structure}
\end{figure}

In figure~\ref{fig:structure}, we display the optimized structures as well as phonons 
dispersion for A0, A60, B0 and B60 systems.
The carbon atoms present in the plane of the HG layers is buckled due to the bonding with 
hydrogen atoms.
A slight lateral displacement of the outer layers is also seen with respect to the central 
layer for A0 and B0 systems.
The layers however approximately retain their stacking angles with respect to the central plane.
Such a behavior has been reported in the 3R-type MoS$_2$ system~\cite{MOS_SOC_SPLITING_2} where 
one of the layers when laterally moved in the plane, showed a second energy minimum.
The optimized lattice constant of A-type and B0 and B60 was found to be 2.50, 2.50, 2.48 and 
2.48 Å, respectively.
Note that the lattice constant for graphene is 2.46 Å. 
Thus, the lattice mismatch between our proposed systems and pristine graphene is less than 
1.5\%, it is important merit for the proposed systems. 


To understand the stability of the systems, phonon dispersion were calculated 
(figure~\ref{fig:structure}.
Most of the phonon frequencies are positive with a few exceptions primarily in the vicinity of 
$\Gamma$-point for most of the systems.
The spread along the \textit{k}-axis of these negative frequencies is also small and likely to 
be due to numerical errors rather than the instability in structure (See 
reference~\cite{negative_phonon_text} for further explanation).
The stability of these systems can further be checked based on its cohesive and formation 
energies.
The energy of formation of trilayer from monolayers and the binding energy per atom can be 
calculated as:

\begin{eqnarray}
E_{\rm formation} &=& E_{\rm trilayer} - \sum E_{\rm layers} \nonumber \\
E_{\rm cohesive}  &=& \frac{E_{\rm trilayer} - \sum E_{\rm atom}}{N_{\rm atoms}} \nonumber
\end{eqnarray}

Negative values of these energies suggest that the system is thermodynamically stable i.e. the 
processes which can decompose and(or) transform these systems are either forbidden and(or) very 
gradual.
In table~\ref{table_binding_energy}, we show the energy of formation of trilayer from 
monolayers and the binding/cohesive energy per atom.
Taking into account the magnitude of the values from the last column of the table, it may be 
concluded that the systems are energetically and thermodynamically favorable.

The negative frequencies near the $\Gamma$ point do not necessarily mean the system is unstable 
as in this case the energetics are favorable.
This has been demonstrated in the work of Mounet et al.~\cite{2D_high_throughput}, where a 
screening of 2D materials from 3D bulk was performed with the help of high throughput 
methodology.
This lead to the creation of a database of 2D materials which includes crystal structure, 
electronic band structure, and phonon dispersion.
They have classified the materials with negative frequencies near the gamma point as stable as 
their energetics were favorable.
Further, as the hydrogenated graphene layer is under strain it will give rise to negative 
frequencies and as mentioned above, it does not imply that the system is unstable.
With these arguments, we may conclude that the systems under consideration are stable.

\begin{table}[H]
	\centering
	\caption{The formation and binding energies of the trilayer systems.}
	\label{table_binding_energy}
	\begin{tabular}{|c|c|c|c|}
		\hline
		\textbf{Name}                                                    & \textbf{Total Energy (Ry)} & \textbf{\begin{tabular}[c]{@{}c@{}}Energy of trilayer\\ formation (Ry)\end{tabular}} & \textbf{\begin{tabular}[c]{@{}c@{}}Binding/Cohesive\\ Energy per atom (Ry)\end{tabular}} \\ \hline
		Graphene                                                         & -36.8942                   & -                                                                                    & -0.6805                                                                                  \\ \hline
		\begin{tabular}[c]{@{}c@{}}Hydrogenated \\ Graphene\end{tabular} & -39.2646                   & -                                                                                    & -0.4737                                                                                  \\ \hline
		A0                                                               & -115.4449                  & -0.0214                                                                              & -0.5172                                                                                  \\ \hline
		A60                                                              & -115.4429                  & -0.0194                                                                              & -0.5170                                                                                  \\ \hline
		B0                                                               & -113.0683                  & -0.0153                                                                              & -0.5790                                                                                  \\ \hline
		B60                                                              & -113.0694                  & -0.0163                                                                              & -0.5792                                                                                  \\ \hline
		\multicolumn{2}{|c|}{Energy of isolated carbon atom}                                          & \multicolumn{2}{c|}{-17.7666 Ry}                                                                                                                                                \\ \hline
		\multicolumn{2}{|c|}{Energy of isolated hydrogen atom}                                        & \multicolumn{2}{c|}{-0.9183 Ry}                                                                                                                                                 \\ \hline
	\end{tabular}
\end{table}

\subsubsection{Electronic Band structure of Proposed Systems}

\begin{figure}[H]
	\centering
	\includegraphics[scale=0.75]{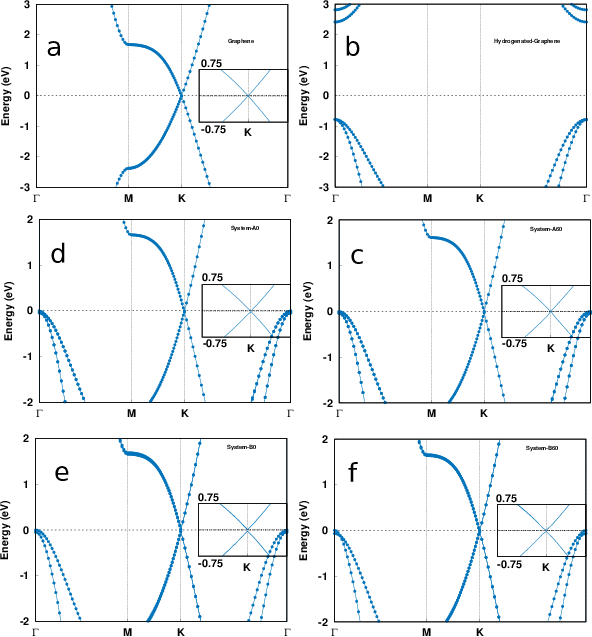}
	\caption{(color online)
		Band structures of (a) pristine graphene and (b) hydrogenated graphene.
		Band structures of the systems (c) A0, (d) A60, (e) B0, and (f) B60.
		Note that the Dirac point is present for the systems A0, A60, B0 and B60 (see inserts) 
		and the band structure of the
		trilayer systems is basically an addition of the band structures of  graphene and the HG 
		band structure.
	}
	\label{band_structure}
\end{figure}

The band structures for all the systems were calculated to assert the feasibility of proposed 
systems to host QSHE.
In figure~\ref{band_structure}, we show the band structures for all the configurations. 
Clearly, from the plot, we can note that for the different configurations of the proposed 
systems, the stacking order or stacking angles seems have little to no effect on the band 
structure of the sandwiched layer.
The band structures of all systems were found to be quite similar, as they consist of the same 
components arranged in different configurations.
From the initial observation, it may be noted that at the \textit{K}-point there is a Dirac 
point similar to that of pristine GR.
However, there is a extremely small and negligible gap, which is less than the error of DFT 
calculations, hence it can be considered as gapless.
There are also some additional bands flanking the Dirac point.
These bands touch the Fermi level, effectively interacting with the Dirac point providing an 
indirect conduction path, indicating these systems to be semimetallic in nature.
If the proposed systems are semimetallic then the possibility of them hosting QSHE is reduced. 
However, detailed analyses of the band structures show that they still hold a possibility of 
hosting QSHE.

For understanding these features further, we plot the \textit{k}-resolved projected density of 
states (PDOS) given in supplementary information (SI) section V.
From these plots, one can observe that the Dirac cone at the \textit{K}-point is formed out of 
the carbon atoms of the GR sheet, with negligible contribution from the HG atoms.
Further, the band structures of the individual pristine GR and HG do not interfere with each 
other indicating a weak chemical coupling between GR and HG.
Hydrogen bonding is observed when the distance of the hydrogen with the heavier and high 
electronegative elements such as O, N, is of the order of 1.5-2.5~\AA.
However, the hydrogen atoms nearest to any of the carbon atoms from the GR layer is about 
2.8~\AA indicating negligible interaction between the HG and GR layers.
In addition, the electron density overlap between these layers is also not present (see SI 
section IV).
Thus, we may be concluded that the presence of the additional band does not affect or alter the 
presence of the  Dirac cone and the origin of the Dirac cone in the system can be attributed to 
the central GR plane, providing a possibility of hosting QSHI state.

In order to examine whether if the HG layer shields multilayer graphene, we studied trilayer 
graphene encapsulated by HG.
Note that even in pristine trilayer graphene the Dirac cone is preserved and has a possibility 
of hosting QSHI states~\cite{QSHE_tri-gr}.
We compute the band structure of this system and found that the characteristics of trilayer 
graphene have been preserved (SI section II) supporting the above-mentioned claim.
Thus, the electronic properties of the graphene layers seem to be protected by the HG layer. 
Recall, that for device applications, the multilayer graphene, which holds QSHE, is generally 
required to be deposited on a substrate.
The substrate states then interact with the graphene system opening up a gap at the Dirac cone 
making it less suitable for QSHE related applications.
One way to protect the graphene states would be to add an HG monolayer between the graphene and 
the substrate.
However, we found that the HG-graphene bilayer itself is very unstable.
Hence, it could be difficult to synthesize and then deposit on a substrate.
On the other hand, our calculations on HG encapsulated monolayer, as well as trilayer graphene, 
indicate that upon deposition on a substrate the graphene states are likely to be protected by 
the HG layers thus retaining the property of graphene layers to host QSHE.

\subsubsection{Effect of Electric Field on Band Structure of Proposed Systems}

\begin{figure}[H]
	\begin{center}
		\includegraphics[scale=0.75]{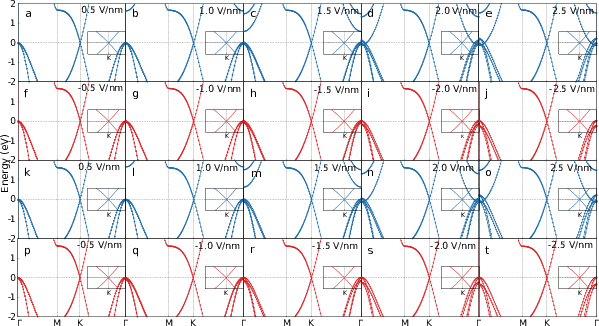}
		\caption{(color online)
			Band structures of system A0~(figures~(a) to (j)) and  A60~(figures~(k) to (t))
			under electric fields of 0.5, 1.0, 1.5, 2.0 and 2.5~V/nm.
			The energy range in the insert is form 0.75 eV to -0.75 eV.
		}
		\label{elec_c_qw}
	\end{center}
\end{figure}

It is well known that the electronic structure of graphene is not affected due to the applied 
electric field.
Hence, an application of an electric field to the HG-GR-HG trilayer should not affect the Dirac 
cone.
In the buckled structure of 2D system such as silicene in the presence of the electric field, 
the potential felt by the sublattice is different which leads to an opening up of a band gap. 
Hence, in the presence of the electric field, if the HG layers due to its buckled nature 
interact with the graphene layer in the presence of an electric field, the Dirac cone may get 
altered.
In order to probe the stability of Dirac cone in presence of the electric field, the HG-GR-HG 
trilayer systems in the A0 and A60 configurations were subjected to electric fields in the range 
from 0.0 to 2.5~V/nm in steps of 0.5~V/nm (see figure~\ref{elec_c_qw}).
The electric field is modeled by a saw-tooth potential which is purely a mathematical technique.
Hence, care must be taken to set the boundary conditions so that the electric field drops to 
zero in the vacuum region away from the HG layers.

\begin{figure}[H]
	\centering
	\includegraphics[scale=0.75]{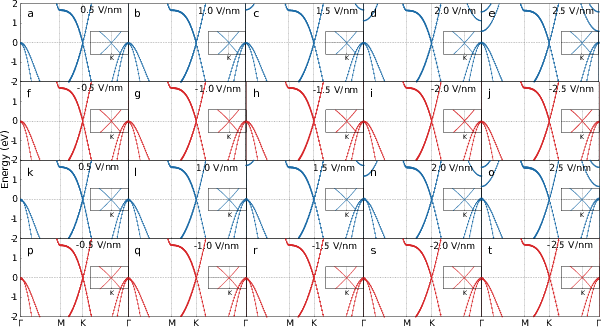}
	\caption{(color online)
		Band structures of system B0~(figures~(a) to (j)) and  B60~(figures~(k) to (t))
		under electric fields of 0.5, 1.0, 1.5, 2.0 and 2.5~V/nm.
		The energy range in the insert is form 0.75 eV to -0.75 eV.
	}
	\label{elec_i_qw}
\end{figure}

Upon switching the electric field, we noted that the Dirac point present at the K-point is not 
affected at any given value of the electric field.
A few bands in the vicinity of the $\Gamma$-point above and below the Fermi level do split.
The projected density of states (PDOS) analyses shows that these bands are from the HG layer 
indicating that the HG states in A0 and A60 configurations do not interfere with the those of GR 
even in the presence of an electric field.
Rather, it acts as a shield for the central GR layer.
A similar conclusion can be made for the systems B0 and B60~(see figure~\ref{elec_i_qw}).
Unlike the A0 and A60 systems, here the Dirac point observed due to the GR sheets, is split 
albeit at high electric fields of about 4 V/nm and higher.
Further, from figure 2, it can be observed that the range of negative phonon dispersion is 
higher when compared to the A0 and A60.
Thus, one may conclude that in the A0 and A60 systems the Dirac point is quite stable. Hence, 
only these systems were considered for further investigation.
From the above observations of the systems A0 and A60 and the derived conclusion, it is clear 
that the effect on the electronic states of GR is not unfavorably affected.
Thus, the HG encapsulated GR holds a possibility of hosting QSHE.

A direct method of testing this claim would be calculating the Z$_2$ topological invariant 
where Z$_2$=1 means that the system is QSHI and Z$_2$=0 means the system is not QSHI.
However, as these systems are constructed via stacking of multiple layers and the system as a 
whole is conducting, the current methods for determining the Z$_2$ invariant cannot be reliably 
used here~\cite{Graphene_Spin_Hall,Z2_without_inversion_symmetry}.
An indirect method of determining if a system can host QSHE is to compute the edge state and to 
count the number of band crossings which should be odd numbered~\cite{Topological_Band_Theory}.
Even this method is difficult to implement since the systems A0 and A60 as a whole have an 
indirect band closing between the bands of GR (at the \textit{K}-point above the Fermi level) 
and HG bands (near the $\Gamma$-point below the Fermi level) which will give the systems 
conducting edges. 
Thus, the only possible \textit{ab-initio} method of determining whether the systems A0 and A60 
can host the QSHE is to verify that the electronic states of GR are not affected.
This has been performed and is concluded that electronic states are indeed not affected.
Hence, the proposed systems A0 and A60 do have a possibility of hosting QSHE.

\subsection{Magnetization in GR - HG heterostructures}

Magnetization in graphene and related materials has been typically studied by adding adatom it 
which can be done using scanning tunneling microscopy method in a controlled 
manner~\cite{MAG_GRAP_1, MAG_GRAP_2}. 
Magnetization may be induced by creating vacancies or distortions in graphene~\cite{MAG_GRAP_3}.
However, such modulation may significantly affect the Dirac cone making graphene lose its QSHI 
states.
Such limitations may be overcome with the proposed HG-graphene system by introducing distortions 
in the HG layer or by forming nanoribbons.
As we shall see below, by both methods, the graphene edge states are observed in addition to 
significantly induced magnetic moments in the heterostructure.

First, we discuss the results of magnetization in nanoribbons.
For this purpose, we constructed nanoribbons with widths up to 20 unit cells by the procedure 
discussed in section~\ref{CM}, in the zigzag and arm-chair conformations of the A0 and A60 
systems and performed DFT calculations.
The systems are further named as R0 and R60 for nanoribbons of system A0 and A60 respectively, 
to distinguish them from their parent systems.

We also added a prefix ``Z'' to denote the zigzag nanoribbon and ``A'' to denote the arm-chair 
nanoribbon.
For example, a zigzag nanoribbon of A0 is denoted as ZR0, while an arm-chair nanoribbon of A60 
is denoted as AR60. See table~\ref{nanoribbons_name} for detailed nomenclature.

\begin{table}[H]
	\centering
	\caption{Labeling schemes for all the nanoribbons.}
	\label{nanoribbons_name}
	\begin{tabular}{c|c|c|}
		\cline{2-3}
		\textbf{}                         & System A0 & System A60 \\ \hline
		\multicolumn{1}{|c|}{Arm - Chair} & AR0       & AR60       \\ \hline
		\multicolumn{1}{|c|}{Zig - Zag}   & ZR0       & ZR60       \\ \hline
	\end{tabular}
\end{table}

Before we discuss the electronic structure, we make note the following characteristics of the 
nanoribbons:
(1) the structures have a unique atomic arrangement, \textit{i.e}, even if zigzag nanoribbons 
are created, one of the edges is in an arm-chair configuration (see 
figure~\ref{ribbion_edge}~(c)),
(2) to observe the distinction between the electronics states of zigzag and arm-chair 
nanoribbons, their dimensions should be as large as possible~\cite{Graphene_nanoribbon}.
The 20 unit cells width nanoribbon was the largest system that we could compute without 
reducing the accuracy of the calculations.
For each the systems, three sets of distinct DFT calculations were performed.
The first set of calculations were performed on systems without any modification where the 
nanoribbons were constructed using the parameters described in section~\ref{CM}.
In the second set of calculations, two unit cells from the center of the top HG layer were 
removed to introduce a defect to increase the number of edges of the HG~(see
figure~\ref{ribbion_edge}~(a)).
The third set of calculations were performed after saturating all the dangling bonds with 
hydrogen for both the systems described above.
Finally, band structures for all the above systems were computed.

\begin{figure}[H]
	\centering
	\includegraphics[scale=0.75]{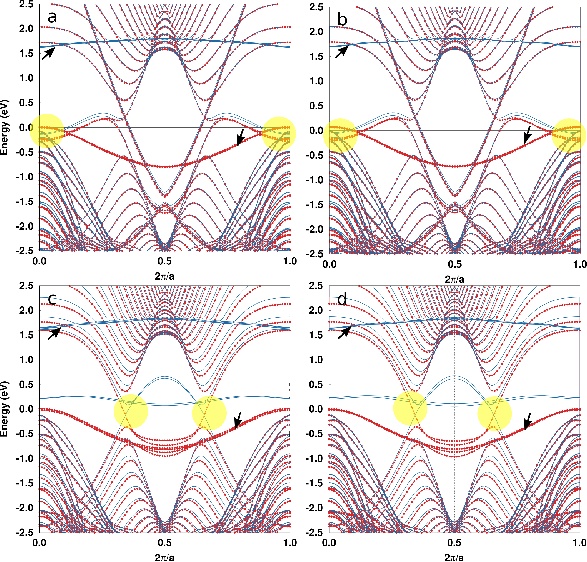}
	\caption{(color online)
		Band structure of (a) ZR0, (b) AR0, (c) ZR60, and (d) AR60.
		The red lines represent up electron bands and blue lines represent down electron bands.
		Some of the important features of the band structure are highlighted by yellow circles.
		The black arrows mark the edge states emerging from the edges of the HG layer.
	}
	\label{band_structure_ribbon}
\end{figure}

The band structure of the system ZR0 is given in figure~\ref{band_structure_ribbon}(a).
As anticipated due to the metallic nature of the systems, there is the presence of metallic 
states at edges.
The number of edge states is even.
However, the edge states are due to contributions from both conduction and valence bands.
For topologically protected states, there should be a pair of bands that connects the 
conduction and valence bands while creating an odd number of Dirac 
crossing/points~\cite{Topological_Insulator_Hassan,Topological_Band_Theory}. 
Thus, one can conclude that these states are not topologically protected.
Interestingly, a presence of absolute magnetization of about 7.18~$\mu_B$/cell is observed, 
causing a considerable spin split in the electronic structure of the system.
This is marked in figure~\ref{band_structure_ribbon} by black arrows.
This marked band originates from the edges of both HG and encapsulated GR layer.
Such behavior is seen in the band structures of all of the systems (AR0, ZR60, and AR60, see 
figure~\ref{band_structure_ribbon}) which are not hydrogenated, asserting the importance of 
passivation of the dangling bonds.
The most remarkable observation is that the electronic structure of the systems is independent 
of the direction of the construction of nanoribbon or the stacking angles between the layers, 
\textit{i.e} the nanoribbons A0 (A60) and R0 (R60) have similar band structures even after 
selecting 20 unit cells.

\begin{figure}
	\includegraphics[scale=0.75]{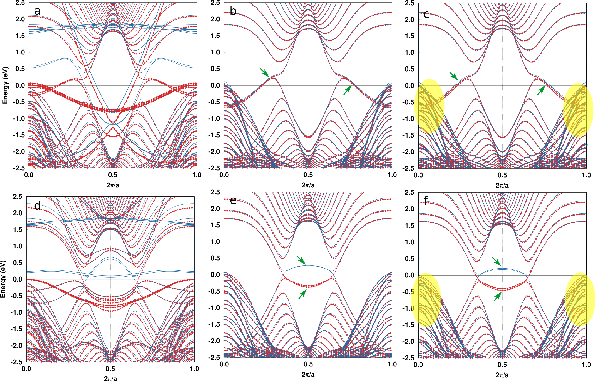}
	\caption{(color online)
		Band structures of (a)~diss-ZR0, (b)~hydrogenated ZR0, (c)~hydrogenated 
		dis-ZR0, (d)~ZR60, (e)~hydrogenated ZR60 and (f) hydrogenated dis-ZR60.
		The red lines represent up electron bands and blue lines represent down 
		electron bands.
		The green arrows point to the edges states emerging from the edge atoms of 
		graphene alone.
		The splitting of bands due to passivation of the HG edges have been marked by
		yellow ovals.
	}
	\label{band_structure_ribbon-2}
\end{figure}

\begin{figure}[H]
	\centering
	\includegraphics[scale=0.75]{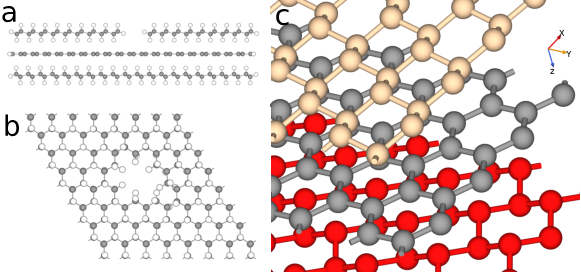}
	\caption{(color online)
		(a) Side view of the nanoribbon with two unit cells removed form the top HG layer, (b) 
		top view of the top HG layer of the system A0 showing the cavity, (c) the edges of the 
		system AR60.
		The hydrogen atoms of the system have been removed.
		The top HG is represented by pinkish/brownish color while the bottom HG is represented using red
		color.
		The graphene layer is colored in grey.
		Note that the edge of graphene layer is armchair whereas that of HG are zigzag.
	}
	\label{ribbion_edge}
\end{figure}

We see features characteristic to the 0\degree and 60\degree stacking angle is present such as 
in the band structure of the ZR0 system (figure~\ref{band_structure_ribbon-2}(a)), in the sense 
that the band crossing occurs at integer value of the \textit{k}-point while for ZR60 system 
(figure~\ref{band_structure_ribbon-2}(c)) the band crossing is observed near the half-integer 
\textit{k}-points.
This has been marked by the translucent yellow circle in figure~\ref{band_structure_ribbon-2}.
The majority of these features arise from the dangling edge bonds which have a very poor 
experimental realization.
Upon hydrogenation, all the ribbons display a significant change in the band structure as well 
as magnetization.
In figure~\ref{band_structure_ribbon-2}(b) and~\ref{band_structure_ribbon-2}(e), we show the 
band structures of the passivated ZR0 and ZR60, respectively.
A substantial change is observed.
The absolute magnetization for both the systems is reduced drastically to about 0.47 (0.71) and 
0.56 (0.69), respectively, due to saturation of the dangling bonds which is reflected well in 
the reduced spin splitting as compared to that of the non-hydrogenated systems.
The interference from HG layers is reduced so significantly that the electronic states near the 
Fermi level are from the graphene sheet only.

It is only logical from this discussion that further calculations can be carried out either on 
zigzag or armchair ribbons.
We carried out the calculations on zigzag ribbons: ZR0 and ZR60.
Further, in order to modulate magnetization, we introduce distortions in these systems by 
removing two unit cells from one of the HG layers.
We label these distorted nanoribbons as ``diss-ZR0 (diss-ZR60)'', respectively.
In figure~\ref{band_structure_ribbon-2}(a) and ~\ref{band_structure_ribbon-2}(c), we show the 
band structures of the diss-ZR0 (non-hydrogenated and 
passivated),~\ref{band_structure_ribbon-2}(d) and~\ref{band_structure_ribbon-2}(f) band 
structures for diss-ZR60 (non-hydrogenated and passivated), respectively.
Due to the presence of an additional HG edge, an increase in the absolute magnetization is 
observed.
It has now increased to 7.49~$\mu_B$/cell for diss-ZR0 further increasing the split in the band 
structure.
As mentioned earlier the major contribution to this split is from the edges of the HG. However, 
for diss-ZR60, the magnetization has decreased by a small value to 7.14~$\mu_B$/cell.
This also explains the decrease in the number of degenerate bands.
Thus, by just increasing the number of edges of HG, the magnetization of the proposed system 
can be altered: an additional parameter for customization of the proposed system.

Interestingly, after hydrogenation, the band structures of distorted ribbons seem to resemble 
those of the undistorted ribbons (compare figure~\ref{band_structure_ribbon-2}(b) for band 
structure of ZR0 with that of diss-ZR0, in figure~\ref{band_structure_ribbon-2}(c)).
However, this observation seems to be misleading as on closer analyzes we found that there are 
some noticeable and important differences.
Firstly, there is a small increase in the spin-split due to increase in edges.
These spin-split states are primarily in the vicinity of the \textit{k}=0 and $2\pi/a$, marked 
by yellow translucent ovals.
This may imply that by fine-tuning the number of edges of HG, one can increase the 
magnetization of the ribbon without significantly interfering the GR states at the Fermi level. 
However, in order to retain most of the features of the graphene states, especially the Dirac 
cone, one may have to increase the width of the nanoribbon up to 50 units cells or 
more~\cite{Graphene_nanoribbon}.

\subsection{Adsorption of Molecules on GR-HG heterostructures}

\begin{figure}[H]
	\centering
	\includegraphics[scale=0.75]{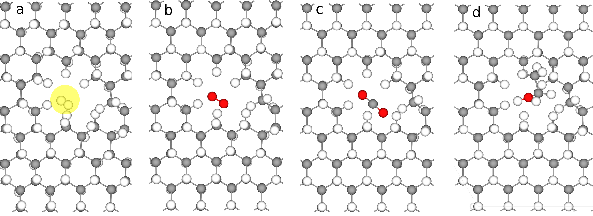}
	\caption{(color online)
		Structure of molecules adsorbed on HG-graphene-HG trilayer system with cavity in the 
		top HG layer.
		Only the top HG layer is shown.
		(a) H$_2$ molecule (in yellow region), (b) O$_2$ molecule, (c) CO$_2$ molecule and 
		(d) ethanol molecule.
	}
	\label{adsorb_molecules}
\end{figure}

Graphene with substitutional non-metal dopants or adatoms has been widely studied for 
applications in sensing molecules and electrocatalysis~\cite{mol_adsorb_1}.
In this section, we test the applicability of the proposed system for adsorption of molecules 
such as H$_2$, O$_2$, carbon dioxide, and ethanol.
As it has been proposed that QSHI like stanene can also be utilized for gas sensing 
applications\cite{Stanene_As_Gas_Absorber}, a brief investigation was performed on the proposed 
system.
Concluded from the above, either by non-passivation of the edges or by increasing the 
distortions in one of the HG layers, magnetization can fine tune.
The magnetization in these systems can certainly be affected by the presence of additional 
factors like adsorbed molecules.
For this purpose, a base system consisting of 8$\times$8$\times$1 number of unit cells of the 
system A0 was created.
In order to induce magnetization, an asymmetric cavity was created on the top HG which was 
hydrogenated, since such a vacancy or impurities is required as these molecules generally do 
not get adsorbed on defect-free surfaces.
In figure~\ref{adsorb_molecules}, we show the image of the top HG with the cavity along with 
the optimized structure of adsorbed H$_2$, O$_2$, CO$_2$ and ethanol molecules on the system A0.
The adsorbed H$_2$ molecule is shown in the yellow translucent circle in 
figure~\ref{adsorb_molecules}a.

\begin{table}[H]
	\centering
	\caption{Magnetization induced due to the presence of cavity and upon adsorption of 
		molecules on to the system A0.}
	\label{magnetization_due_to_adsorption}
	\begin{tabular}{|l|c|}
		\hline
		~ \textbf{Name}   & \textbf{\begin{tabular}[c]{@{}c@{}}Absolute Magnetization\\ ($\mu_{B}/cell$)\end{tabular}}\\ \hline
		~ System A0       & 0.1                                                                                       \\ \hline
		~ A0 : H$_2$      & 2.35                                                                                      \\ \hline
		~ A0 : O$_2$      & 7.23                                                                                      \\ \hline
		~ A0 : CO$_2$     & 7.28                                                                                      \\ \hline
		~ A0 : ethanol  ~ & 7.29                                                                                      \\ \hline
	\end{tabular}
\end{table}

Geometry optimization of the distorted A0 system with the molecules adsorbed was carried out. 
These molecules are likely to change the magnetization upon adsorption on the A0 surface. 
However, the total number of the atoms exceeds 600 making the geometry optimization within DFT 
framework difficult.
Hence, we performed the geometry optimization using the Brenner force field for the substrate 
distorted A0 system and RelaxFF force field for the adsorbed molecules as implemented in ATK 
VNL package~\cite{VNL-2017} until the maximum absolute forces on the atoms were less than 
$10^{-4}$~eV/\AA.
Then the optimized geometry was used to calculate magnetization using DFT.
The DFT parameters are kept the same as described in section~\ref{CM}, except that the 
calculations were performed for $\Gamma$-point only due to a large number of atoms.
We observed a distinct and significant change in the magnetization of the system for each 
molecule as summarized in Table~\ref{magnetization_due_to_adsorption}.
These changes can be easily detected in the experiments. Hence, we may conclude that the 
proposed system can be used to detect these molecules.

\section{Conclusion}

We performed DFT calculations on hydrogenated graphene - graphene trilayer systems with PBE 
exchange-correlation potential including the DFT-D2 Van der Waals correction for representing 
accurately the phonon dispersion effects.
We found that the heterostructure of graphene encapsulated by hydrogenated graphene could be a 
promising candidate for a variety of device applications.
A detailed investigation shows that this system does have a possibility of preserving and 
protecting the edge states of GR.
The presence of the HG layer makes the system magnetic which can further be tuned by 
controlling the vacancies in HG layers.
These defects or vacancies in the outermost HG layer can also be used to detect molecules such 
as H$_2$, O$_2$, CO$_2$ and ethanol since upon adsorption a significant change in magnetic 
moments is observed which can be detected experimentally.
An additional important feature that was noted was the system independence on the direction of 
nanoribbons providing an additional simplicity for experimental realization of the proposed 
system.
Our results on trilayer graphene encapsulated by HG indicate that the above observations can be 
extended to these systems with a larger number of central graphene layers.
Interestingly, such systems are relatively simpler to fabricate by first exfoliating or 
synthesizing multilayer graphene and then hydrogenating the outermost layers.
This work brings out the versatility of the multilayer GR sandwiched between two HG layers and 
thus has the merit of being experimentally investigated.

\begin{acknowledgement}
The computational work described here is performed at the High-Performance Computational 
Facility at IUAC, New Delhi, India.
We would like to express our gratitude to them.
Also, we would like to thank the University Grants Commission of India for providing partial 
funding for the research work through the UGC-BSR Research Startup Grant (Ref. 
No.F.30-309/2016(BSR)).
\end{acknowledgement}

\begin{suppinfo}
See supplementary information for the crystal structure data, band structures of HG encapsulated
bilayer and trilayer graphene, $k$-resolved projected density of states~(PDOS) and charge
densities.
	\begin{itemize}
		\item Section I		: Coordinates of Proposed Systems
		\item Section II	: Band structure of Trilayer graphene encapsulated with HG
		\item Section III	: \textit{k}-resolved Projected Density of States
		\item Section IV	: Charge Density
		\item Section V		: Selected \textit{k}-resolved PDOS of nanoribbon
		\item Section VI	: Bilayer Structures
		\item Section VII	: Band structure of unstable system
	\end{itemize}
	
\end{suppinfo}

\bibliographystyle{achemso-demo}

\begin{thebibliography}{10}
\bibitem[neg()]{negative_phonon_text}
	Generally, presence of negative frequencies indicate a structural instability.
	However, in this case, a careful examination of the phonon frequencies is
	required before coming to a conclusion. The magnitude of negative phonon
	frequencies is not significant and so is it's spread in momentum space,
	$i.e.$ they are localized near the $\Gamma$-point only, for most of the
	systems. We have increased various input parameters such as the cutoff
	radius, number of $k$-points for optimization, and found it to reduce the
	range of $k$-points with negative frequencies. Thus, these small magnitude
	and spread of negative frequencies are likely due to numerical instabilities
	or errors rather than the structural instability. In addition, we also
	observed that the change in the electronic properties are negligible when
	compaired with the structure having a higher negative phonon dispersion.
	Further: these negative frequencies are between two successive special
	$k$-points within the Brillouin zone, which generally raises due to out-of
	plane buckling and/or due to the strain due to lattice mismatch, as explain
	in text. Here these negative frequencies are very weak comparison, hence upon
	fabrication on a substrate they are likely to die out.\relax
	\mciteBstWouldAddEndPunctfalse
	\mciteSetBstMidEndSepPunct{\mcitedefaultmidpunct}
	{}{\mcitedefaultseppunct}\relax
	\EndOfBibitem
\end{thebibliography}

\end{document}